\documentclass[aps,prl,reprint,superscriptaddress,longbibliography,floatfix]{revtex4-2}
\usepackage{array}[=2016-10-06]
\usepackage{amsmath,amssymb,bm,graphicx,booktabs}
\usepackage[pass]{geometry}
\usepackage{microtype}\AtBeginDocument{\microtypesetup{activate=false}}
\usepackage{placeins}
\usepackage[hypertexnames=false,colorlinks=true,linkcolor=blue,citecolor=blue,urlcolor=blue]{hyperref}
\graphicspath{{figures/}}
\newcommand{\GeV}{\mathrm{GeV}}
\newcommand{\keV}{\mathrm{keV}}
\newcommand{\snu}{\widetilde\nu}
\newcommand{\dd}{\mathrm d}
\newcommand{\hc}{\mathrm{h.c.}}
\newcommand{\eV}{\mathrm{eV}}
\newcommand{\smref}[2]{\hyperlink{supp-sm-#1}{Suppl., #2}}
\newcommand{\prlsection}[1]{\par\noindent\hspace*{\parindent}\textit{#1}---}
\begin{document}

\title{Explaining the LZ High-Energy Recoil Event with Inelastic Sneutrino Dark Matter in Supersymmetry}
\author{Jingwei Lian}
\email{lianjw@hist.edu.cn}
\affiliation{Henan Institute of Science and Technology, Xinxiang 453003, P. R. China}
\affiliation{Centre for Theoretical Physics, Henan Normal University, Xinxiang 453007, P. R. China}
\author{Jin Min Yang}
\email{jmyang@itp.ac.cn}
\affiliation{Centre for Theoretical Physics, Henan Normal University, Xinxiang 453007, P. R. China}
\affiliation{Institute of Theoretical Physics, Chinese Academy of Sciences, Beijing 100190, P. R. China}
\begin{abstract}
We study an inelastic sneutrino dark matter origin of the LZ high-energy event in the inverse-seesaw next-to-minimal supersymmetric standard model. Approximate lepton-number conservation protects the small mass splitting, while singlet interactions separate freeze-out from solar-neutrino production. The LZ recoil and solar constraints overlap even for a fully cooled population in equilibrium. Three thermal mechanisms reproduce the relic abundance and satisfy all relevant constraints. Photon-dominated annihilation produces gamma-ray boxes, offering a complementary test of this interpretation.
\end{abstract}
\maketitle

\prlsection{Introduction}
The LUX-ZEPLIN (LZ) collaboration recently reported a nuclear-recoil event at $248\pm23\,\mathrm{(stat)}\pm23\,\mathrm{(sys)}\,\keV$ in an exposure of $2.84$ tonne-years, with a maximum local significance of $3.4\sigma$ and a global significance of $2.6\sigma$~\cite{LZHE}. The interpretation via endothermic dark-matter (DM) inelastic scattering selects the high-speed halo population and suppresses low-energy recoils~\cite{InelasticDM}. Solar acceleration enables capture through the same interaction, linking this interpretation to solar-neutrino searches~\cite{IceCubeInelastic,Pospelov2026,SolarTests2026}.

A nearly pure thermal Higgsino near $1.1\,\mathrm{TeV}$ offers a supersymmetric interpretation~\cite{Freese2026,HiggsinoHalo}. However, a neutral-state splitting of a few hundred keV requires very heavy gauginos or cancellations in the minimal supersymmetric standard model. Electroweak annihilation channels create strong tension with solar-neutrino limits for a cooled population~\cite{Freese2026,HiggsinoHalo,Pospelov2026}, while high-energy xenon sidebands impose further constraints~\cite{Rodd2026}. This requires extremely fine-tuned
interactions controlling the splitting, scattering rate and neutrino yield per annihilation.

In the next-to-minimal supersymmetric standard model (NMSSM), the singlet can raise the tree-level Higgs mass, reducing the stop contribution needed for $125\,\GeV$, and generate an effective $\mu$ parameter through its vacuum expectation value~\cite{NMSSMReview}. A conventional type-I seesaw extension faces two obstacles. For comparable weak-scale soft parameters, heavy-singlet decoupling leaves a sneutrino splitting of order the light-neutrino mass~\cite{Minkowski1977,GrossmanHaber}. At the TeV scale, reproducing neutrino masses without cancellations requires $Y_\nu\sim10^{-6}$. For trilinear couplings proportional to $Y_\nu$ and nondegenerate scalar masses, this suppresses the active sneutrino component and inelastic $Z$-exchange signal~\cite{Yaguna2008}.

In the inverse seesaw, small lepton-number violation permits sizable Yukawa couplings~\cite{InverseSeesaw,DeRomeriHirsch2012}. Earlier work explored inelastic sneutrinos in gauged inverse seesaw models in connection with CRESST-II~\cite{An2012Sneutrino}. 
Related singlet-doublet fermion models link pseudo-Dirac dark matter to radiative neutrino masses~\cite{Konar2020PseudoDirac}, with extensions to leptogenesis in nonstandard cosmologies~\cite{Konar2021Leptogenesis}.
The inverse-seesaw NMSSM (ISS-NMSSM) supports singlet sneutrino dark matter~\cite{Cao2021gminus2,Cao2020gminus2,Cao2020,Cao2017,Kang2016}. 
Here we interpret the LZ event within this framework.
Our benchmarks fit neutrino data, protect the splitting through approximate lepton-number conservation, and separate freeze-out from solar-neutrino production through singlet interactions. 
For a standard halo and a fully cooled population in solar capture--annihilation equilibrium, the LZ recoil and solar capture rates determine the required suppression of the neutrino yield. Dalitz decays produce neutrinos even when diphoton decays dominate, linking solar tests to Fermi-LAT dwarf-galaxy constraints~\cite{FermiDwarfs14} and gamma-ray box searches with the Cherenkov Telescope Array Observatory (CTAO)~\cite{GammaBoxes2012,CTAOProd3b}. 

\prlsection{Symmetry protection of the inelastic splitting}
For three singlet pairs with lepton numbers $L(\hat N)=-1$ and $L(\hat X)=+1$, the relevant interactions are
\small
\begin{eqnarray}
& &W=W_{\rm NMSSM}+Y_\nu\hat L\hat H_u\hat N+\lambda_N\hat N\hat S\hat X+\tfrac12\mu_X\hat X\hat X,\\
& & V_{\rm soft}\supset\tfrac12b_X\widetilde X\widetilde X+\hc .
\label{eq:W}
\end{eqnarray}
\normalsize 
The symmetric matrices $\mu_X$ and $b_X$ have mass dimensions one and two, respectively, and independently violate lepton number. With $m_D=Y_\nu\langle H_u^0\rangle$ and $M_R=\lambda_N\langle S\rangle$,
\begin{equation}
 m_\nu^{\rm tree}\simeq m_D(M_R^T)^{-1}\mu_XM_R^{-1}m_D^T .\label{eq:mnu}
\end{equation}
The heavy singlets form quasi-Dirac pairs with masses set by $M_R$. For fixed $m_D$ and $M_R$, the light-neutrino masses vanish at tree level as $\mu_X\to0$, linking their smallness to approximate lepton-number conservation~\cite{InverseSeesaw}.
We assume real parameters and conserved $R$-parity, denoting the stable ground state and its inelastic partner by $\snu\equiv\snu_1^R$ and $\snu'\equiv\snu_1^I$. In a one-generation approximation, let $(V_L,V_N,V_X)$ be the normalized components of the light eigenstate in the basis $(\snu_L,\widetilde N^*,\widetilde X)$ in the lepton-number-conserving limit. When the CP-even state is lighter, the leading mass splitting is
\begin{equation}
 \delta=m_{\snu'}-m_{\snu}\simeq
 \frac{|2M_R\mu_XV_NV_X+b_XV_X^2|}{m_{\snu}}.\label{eq:gap}
\end{equation}
When $\mu_X=b_X=0$, the two CP components recombine into a complex scalar carrying conserved lepton number. Splitting them requires a $\Delta L=2$ mass insertion. Gauge, Higgs and soft interactions that conserve lepton number can shift the common mass but cannot lift this degeneracy, to all orders in perturbation theory. Radiative contributions to the splitting must therefore contain $\mu_X$ or $b_X$~\cite{GrossmanHaber,SneutrinoLNV,Hirsch1997BL}. Approximate lepton-number conservation makes a gap far below the weak scale natural when the breaking parameters are individually small.

The terrestrial signal arises from endothermic inelastic scattering on a nucleus $A$,
\begin{equation}
 \snu+A\ \xrightarrow{\ Z\ }\ \snu'+A,\qquad \delta>0 .\label{eq:scattering}
\end{equation}
The effective transition coupling $k_Z=\sum_{\alpha=e,\mu,\tau}V^R_{1\alpha}V^I_{1\alpha}$ gives a scattering rate proportional to $k_Z^2$. We also denote the ground-state active fraction by
$f_L=\sum_\alpha|V^R_{1\alpha}|^2$. A real scalar has no diagonal $Z$ current. Elastic spin-independent (SI) scattering through Higgs exchange is included with the relative signs of the proton and neutron amplitudes. A splitting of $300$--$330\,\keV$ selects the high-speed halo population at Earth, whereas solar acceleration enlarges the phase space for capture. The Supplemental Material (Suppl.)~\cite{SupplementalMaterial} gives the mass matrices and numerical methods.

\prlsection{Recoil and solar feasibility}
For nuclear mass $m_A$ and reduced mass $\mu_A$, endothermic recoil requires
\small
\begin{equation}
 v_{\min}(E_R)=\frac{m_AE_R/\mu_A+\delta}{\sqrt{2m_AE_R}},~~~
 v_*\equiv\min_{E_R}v_{\min}=\sqrt{\frac{2\delta}{\mu_A}} .\label{eq:vmin}
\end{equation}
\normalsize
The differential rate scales as $\dd R_Z/\dd E_R\propto k_Z^2F_A^2\eta_v(v_{\min})$, where $F_A$ is the Helm form factor and $\eta_v$ is the mean inverse speed in the laboratory frame. For $^{131}$Xe, the selected parameter points give $v_*\simeq705$--$788\,\mathrm{km\,s^{-1}}$, compared with $735$--$764\,\mathrm{km\,s^{-1}}$ for the three benchmarks. We adopt a truncated Maxwellian standard halo model (SHM) with $\rho_{\snu}=0.4\,\GeV\,\mathrm{cm}^{-3}$. The inelastic and elastic rates are folded with the xenon efficiency and a Gaussian energy resolution of $32.53\,\keV$. We denote the reconstructed signal count density by $g(E)$
and its integral over $50$--$300\,\keV$ by $s$.
The observed event energy is $E_*=248\,\keV$.
To penalize an excessive high-energy recoil tail, we
supplement the energy-only statistic $q$ with a
sideband (SB) penalty:
\begin{equation}
 q=-2\ln\frac{\mathcal L_{\rm prof}}{\mathcal L_{\rm ref}},
 \qquad q_{\rm SB}=q+2N_{\rm SB}.
 \label{eq:recoilq}
\end{equation}
Here $\mathcal L_{\rm prof}=\max_{b\ge0}\mathcal L$
profiles the background expectation $b$, and
$\mathcal L_{\rm ref}$ is a fixed reference common to
all parameter points. The sideband count $N_{\rm SB}$
is predicted in the true-recoil interval
$350$--$590\,\keV$ before detector acceptance;
$2N_{\rm SB}$ is the zero-event Poisson penalty
assuming unit sideband acceptance~\cite{Rodd2026}.
We select points with $q_{\rm SB}<2.3$. This criterion does not define an experimental confidence region, and its response assumptions are given in \smref{recoil}{Sec.~S4}.

For pure $Z$-exchange and identical active components of the CP partners ($V^R_{1\alpha}=V^I_{1\alpha}$, hence $k_Z=f_L$), the accepted LZ signal count $N_{\rm LZ}=f_L^2N_0$ and solar capture rate $C_\odot=f_L^2C_0$ have the same coupling dependence. Here $N_0$ and $C_0$ are evaluated at unit coupling for each $(m_{\snu},\delta)$. Capture--annihilation equilibrium gives $\Gamma_{\rm ann}=C_\odot/2$. For a fully cooled population, define $k_\nu=\mathcal Y_\nu/\mathcal Y_{WW}$ as the ratio of detector-weighted neutrino yields per annihilation. The solar signal-to-limit ratio is then
\begin{equation}
 R_\odot=\frac{C_\odot k_\nu}{2\Gamma_{WW}^{90}}
 =N_{\rm LZ}\frac{C_0}{N_0}\frac{k_\nu}{2\Gamma_{WW}^{90}}.\label{eq:bridge}
\end{equation}
The reference rate $\Gamma_{WW}^{90}$ is inferred from the published IceCube limit for $W^+W^-$ annihilation at the same DM mass, using the published capture-to-annihilation conversion~\cite{IceCubeTenYear,HiggsinoHalo}. Reducing the active fraction suppresses both capture and nuclear recoils, whereas reducing the neutrino yield weakens only the solar signal. The fully cooled equilibrium assumption therefore tests the particle-physics mechanism without invoking delayed cooling or incomplete equilibration.

\begin{figure*}[t]
\centering\includegraphics[width=\textwidth]{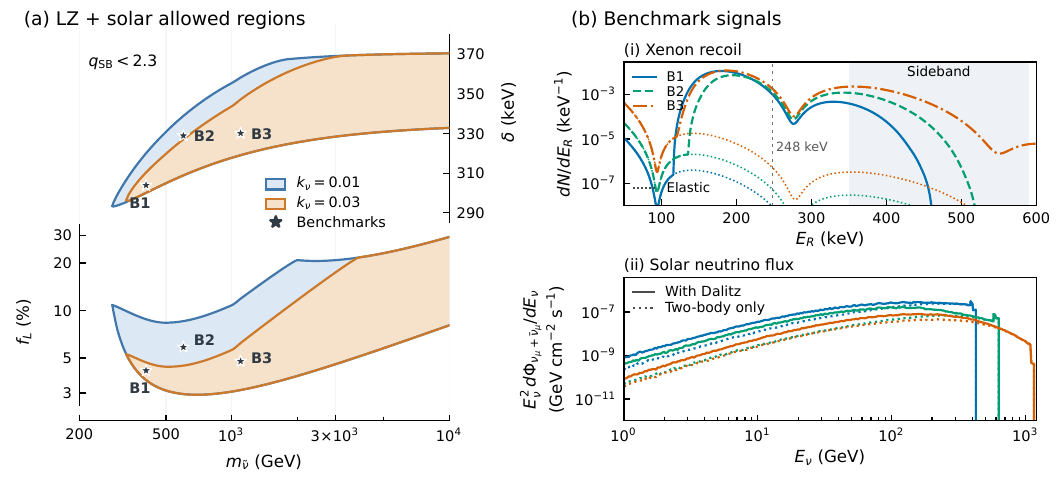}
\vspace*{-1.0cm}
\caption{(a) Parameter regions satisfying $q_{\rm SB}<2.3$ and $R_\odot\leq1$ for pure $Z$ exchange with $k_Z=f_L$, assuming a fully cooled population in capture--annihilation equilibrium. The response-weighted neutrino-yield ratio is $k_\nu=0.01$ (blue) or $0.03$ (orange). The upper and lower bands are separate projections onto $\delta$ (right axis) and $f_L$ (left axis), so arbitrary pairings need not satisfy both criteria. Stars mark the benchmarks, whose neutrino yields are evaluated individually. (b)(i) Xenon spectra for $2.84$ tonne-years, including inelastic $Z$ exchange and elastic Higgs exchange. Thin dotted curves show the elastic contribution. Shading marks the $350$--$590\,\keV$ sideband entering $q_{\rm SB}=q+2N_{\rm SB}$. The vertical line marks $248\,\keV$. (ii) Solar $\nu_\mu+\bar\nu_\mu$ fluxes at Earth before propagation through terrestrial matter, with annihilation rates set by total elastic and inelastic capture. Surviving neutrino lines are averaged over $\Delta\log_{10}E=0.04$ for display. Solid and dotted curves include and omit conversion decays, respectively. Colors identify the benchmarks.}\label{fig:solar}
\end{figure*}

Figure~\ref{fig:solar}(a) shows the interplay between recoil and solar constraints. The lower mass limits are approximately $284$ and $329\,\GeV$ for $k_\nu=0.01$ and $0.03$, respectively. The lower boundary in $\delta$ selects spectra compatible with the $248\,\keV$ candidate. Increasing $\delta$ reduces the accessible halo population until the mixing required by LZ reaches the solar upper bound. At higher masses, recoil and sideband requirements set the upper boundaries. The $f_L$ envelopes change slope near $2$ and $4\,\mathrm{TeV}$ and converge when the solar constraint becomes subdominant. The lower limit on $f_L$ reaches approximately $3\%$ near $0.7\,\mathrm{TeV}$ and then rises as the DM number density decreases. For $m_{\snu}\gg m_{\rm Xe}$, the reduced mass approaches $m_{\rm Xe}$, the splitting interval approaches $333$--$371\,\keV$, and the active fraction needed for a fixed event count scales approximately as $\sqrt{m_{\snu}}$. The bands extend to the scan boundary at $10\,\mathrm{TeV}$. The benchmarks also satisfy the relic-density and neutrino requirements.

\begin{table}[t]
\centering\small
\caption{Properties of the three benchmarks. The active fraction is $f_L=\sum_\alpha|V^R_{1\alpha}|^2$. $N_{a-b}$ denotes the reconstructed count in $[a,b]\,\keV$ for $2.84$ tonne-years. $N_{\rm SB}$ is the predicted true-recoil count in $350$--$590\,\keV$ before detector acceptance, entering $q_{\rm SB}=q+2N_{\rm SB}$. The ten-year solar signal-to-limit ratio $R_\odot$ spans three flavor prescriptions and uses log-linear interpolation of the $W^+W^-$ reference limit.}\label{tab:key}
\setlength{\tabcolsep}{5pt}
\begin{tabular}{lrrr}\toprule
Quantity & B1 & B2 & B3\\\midrule
$m_{\snu}$ (GeV) & 405.00 & 600.94 & 1100.00\\
$\delta$ (keV) & 303.92 & 328.97 & 330.18\\
$f_L$ (\%) & 4.161 & 5.854 & 4.765\\
$N_{50-300}$ & 0.753 & 0.447 & 0.834\\
$N_{210-270}$ & 0.195 & 0.160 & 0.247\\
$N_{\rm SB}$ & 0.0114 & 0.0592 & 0.1673\\
$q_{\rm SB}$ & 2.180 & 1.874 & 2.078\\
$R_\odot$ & 0.789--0.863 & 0.747--0.853 & 0.752--0.811\\
\bottomrule
\end{tabular}
\end{table}

Figure~\ref{fig:solar}(b) includes elastic Higgs exchange in both the recoil and capture rates. This contributes at most $2.85\%$ of the accepted recoil count and $6.64\times10^{-6}$ of the capture rate, so inelastic scattering dominates. Our analysis of the public IceTracks-DR2 sample~\cite{IceTracksDR2,IceTracksDR2Data} does not exclude the three benchmarks for the reference detector response.

Infall of the Large Magellanic Cloud (LMC) can enhance the high-speed halo population~\cite{BeslaLMC2019}, whose properties vary substantially among simulations~\cite{OHareHighVelocity2026}. Figure~S2 compares the Maxwellian distribution with the Halo~13 MW+LMC template~\cite{SmithOrlikLMC}. A larger high-speed population increases both signal and sideband counts, so nuclear-response and halo assumptions affect the selected parameter points. Although joint recoil--solar compatibility is demonstrated for SHM--Helm with a fully cooled population in equilibrium, viable model realizations persist for the other nuclear-response and halo templates considered.

Away from equilibrium, $R_\odot\simeq F_\odot C_\odot k_\nu/(2\Gamma_{WW}^{90})$, where $F_\odot=2\Gamma_{\rm ann}/C_\odot$ and $k_\nu$ is evaluated for the spatial annihilation distribution. Exploratory coannihilation and thermal-resonance examples give $F_\odot=4.45\times10^{-4}$ and $5.52\times10^{-3}$, respectively, with low-velocity annihilation dominated by $W^+W^-$ and $ZZ$ (\smref{nonequilibrium}{Sec.~S5}). Their small annihilation rates also place the Galactic photon signals orders of magnitude below the assumed CTAO sensitivity. We therefore focus on equilibrium benchmarks with correlated solar-neutrino and gamma-ray signals.

\prlsection{Photon-dominated annihilation and Dalitz decays}
Singlet interactions allow $\snu\snu\to a_sa_s$, where $a_s$ is a predominantly singlet CP-odd Higgs boson. Its small doublet component suppresses tree-level decays into fermions, allowing charged-particle loops to make $a_s\to\gamma\gamma$ dominant~\cite{DiphotonPseudoscalar}. In equilibrium, the neutrino yield per annihilation controls the solar signal at fixed capture rate.

The same photon coupling induces Dalitz decays $a_s\to\gamma\gamma^*\to\gamma\ell^+\ell^-$ and hadronic conversions, with additional contributions from a virtual $Z$. We include photon--$Z$ interference, three-body kinematics and tau polarization. At low invariant mass, hadronic final states are described using exclusive electromagnetic channels and vector resonances, matched to a quark continuum at higher invariant mass~\cite{Hazma2022,PDGHadronic}. Light hadrons stop in the Sun, while heavy-flavor and tau decays contribute to the high-energy neutrino spectrum. Normalizing the two-body and conversion modes to their combined width reduces the diphoton branching fractions from $0.9981$, $0.9994$ and $0.9915$ to $0.9127$, $0.9135$ and $0.9099$ for B1--B3. The box photon yield consequently decreases by $8$--$9\%$. The additional decay modes produce high-energy neutrinos to which the detector is particularly sensitive. The solar and gamma-ray calculations use the same branching fractions and hadronic matching prescription (\smref{decays}{Sec.~S5}).

For B2, including conversion decays while keeping the solar capture rate and mixed-flavor detector response unchanged raises $k_\nu$ from $0.01106$ to $0.01676$ (52\%). The corresponding solar signal-to-limit ratio increases from $0.563$ to $0.853$, as illustrated in Fig.~\ref{fig:solar}(b)(ii). Although mediator masses of $64$--$76\,\GeV$ forbid on-shell $Z\gamma$, $W^+W^-$ and $ZZ$ decays, virtual conversions and direct annihilation into electroweak final states still produce neutrinos. Solar searches are therefore sensitive to the small nonphotonic branching fractions, while most of the annihilation energy is emitted as gamma rays.

\prlsection{Thermal mechanisms and gamma-ray tests}
Freeze-out depends on the annihilation velocity dependence and nearby superpartner abundances, whereas cooled solar annihilation involves the ground-state pair. Three thermal mechanisms reproduce the relic abundance while retaining photon-dominated annihilation at low velocities:

\textit{(i) Thermal resonance.} In B1, annihilation proceeds through $\snu\snu\to h_s^*\to a_sa_s$, where the singlet-dominated CP-even Higgs boson has $m_{h_s}\simeq857.25\,\GeV$, approximately $47.25\,\GeV$ above the ground-state pair threshold. Thermal motion accesses the resonance during freeze-out, while the cooled solar population lies far below it, separating thermal and low-velocity annihilation rates.

\begin{figure}[!t]
\centering\includegraphics[width=\columnwidth]{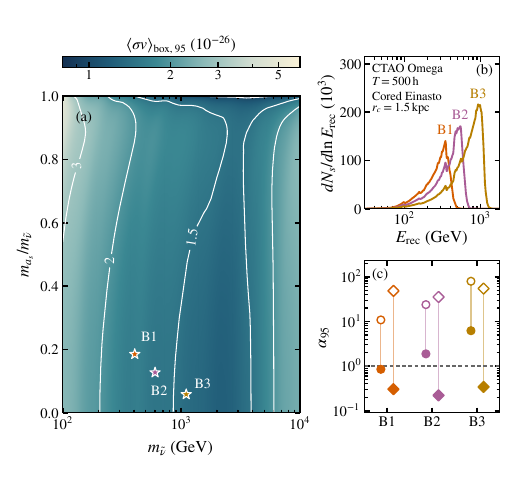}
\vspace*{-1.0cm}
\caption{Gamma-ray comparisons and projected sensitivity for $500$ h of CTAO observations. (a) Sensitivity to a pure box spectrum, expressed as $\langle\sigma v\rangle_{{\rm box},95}$ in units of $10^{-26}\,\mathrm{cm^3\,s^{-1}}$, for an Einasto profile with a $1.5$ kpc core. Stars mark the benchmark mass ratios. (b) Predicted reconstructed counts per logarithmic energy interval, including Dalitz decays, for the same profile. (c) Model-to-limit ratios $\alpha_{95}=\langle\sigma v\rangle_{\rm model}/\langle\sigma v\rangle_{95}$. Circles show expected CTAO limits and diamonds the approximate H.E.S.S. comparison~\cite{HESSIGS2022}. Open and filled symbols correspond to core radii of zero and $1.5$ kpc, respectively. Unity marks the sensitivity threshold. The H.E.S.S. result depends strongly on the ON-minus-OFF signal contrast and is used only to illustrate halo-profile dependence.}\label{fig:gamma}
\end{figure}

\textit{(ii) Electroweakino coannihilation.} In B2, nearly degenerate neutralinos and charginos remain abundant at freeze-out and annihilate into Standard Model (SM) particles through processes such as $\widetilde\chi_1^0\widetilde\chi_1^\pm\to\mathrm{SM}$ and $\widetilde\chi_1^+\widetilde\chi_1^-\to\mathrm{SM}$. Electroweakino initial states account for at least $60.48\%$ of thermal depletion, while $a_sa_s$ final states contribute at least $2.62\%$. By contrast, $99.08\%$ of low-velocity ground-state annihilation proceeds into $a_sa_s$. Neutralino and chargino mass differences from the sneutrino of $0.323$ and $1.765\,\GeV$ sustain coannihilation and motivate collider searches for compressed spectra.

\textit{(iii) Singlet-pair annihilation.} In B3, $\snu\snu,\snu'\snu'\to a_sa_s$ account for at least $87.6\%$ of thermal depletion through interfering contact and exchange amplitudes. Both nearly degenerate CP partners contribute at freeze-out, whereas ground-state annihilation in the Sun has a $97.9\%$ branching fraction into $a_sa_s$.

All three mechanisms also allow Galactic gamma rays from $\snu\snu\to a_sa_s\to4\gamma$. The singlet vacuum expectation value and couplings determine the Higgs spectrum, Higgsino masses and sneutrino interactions~\cite{Cao2017,NMSSMReview}, while Higgs mixing and charged-particle loops control the diphoton branching fraction~\cite{DiphotonPseudoscalar}. Squark masses enter through Higgs radiative corrections and are included consistently. These parameters correlate freeze-out, nuclear scattering and photon production. Boosted diphoton decays produce a box-shaped spectrum with endpoints determined by $m_{\snu}$ and $m_{a_s}$~\cite{GammaBoxes2012}. The predicted detector spectra include both this box and photons from conversion decays.

Figure~\ref{fig:gamma}(a) illustrates the dependence on the mediator mass. As $r=m_{a_s}/m_{\snu}$ approaches unity, the box narrows toward a line at $m_{\snu}/2$, changing its contrast with the detector background. The small mass ratios of B1--B3 give broad boxes with upper endpoints close to $m_{\snu}$. Their reconstructed spectra in panel (b) extend to different energies, relating the DM mass inferred from recoils to the gamma-ray endpoint.

Defining $\alpha_{95}=\langle\sigma v\rangle_{\rm model}/\langle\sigma v\rangle_{95}$, the $500$ h CTAO forecast for an Einasto profile with a $1.5$ kpc core gives $0.849$, $1.880$ and $6.201$ for B1--B3. B2 and B3 lie within the projected sensitivity, while B1 is close to the boundary. Removing the core strengthens the sensitivity, as shown in panel (c). These projections depend on the halo profile, background model and instrument response~\cite{CTAOProd3b}, specified in \smref{photons}{Sec.~S6}.

\begin{table}[!t]
\centering\small
\caption{Experimental probes of B1--B3 and the physical quantities constrained. The solar comparison assumes fully cooled capture--annihilation equilibrium. Numerical results and analysis methods are given in the Supplemental Material.}\label{tab:constraints}
\setlength{\tabcolsep}{3pt}
\renewcommand{\arraystretch}{1.12}
\begin{tabular}{@{}>{\raggedright\arraybackslash}p{0.25\columnwidth}>{\raggedright\arraybackslash}p{0.70\columnwidth}@{}}\toprule
Probe & Experimental input and physical test\\\midrule
Relic abundance & Planck: thermal dark-matter density, with $20\%$ theory tolerance~\cite{Planck2018}.\\
\addlinespace[2pt]
Nuclear recoils & LZ: high-energy inelastic recoil spectrum, sideband counts and low-energy elastic SI scattering~\cite{LZHE,LZ2024}.\\
\addlinespace[2pt]
Solar neutrinos & IceCube IC79, ten-year solar search and fourteen-year IceTracks-DR2: capture and neutrino yield~\cite{nulike,IC79Data,IceCubeTenYear,IceTracksDR2,IceTracksDR2Data}.\\
\addlinespace[2pt]
Gamma rays & Fermi-LAT dwarf-galaxy limits, an approximate H.E.S.S. Galactic-center comparison, and projected CTAO sensitivity to gamma-ray boxes~\cite{FermiDwarfs14,HESSIGS2022,CTAOProd3b}.\\
\addlinespace[2pt]
Neutrino oscillations & NuFIT~6.0 mass splittings and mixing angles, with marginal nonunitarity bounds from electroweak and flavor data~\cite{NuFIT6,Nonunitarity}.\\
\addlinespace[2pt]
Higgs sector & LEP, Tevatron and LHC exclusions from HiggsBounds, and observed Higgs mass and signal strengths from HiggsSignals~\cite{HiggsBounds5,HiggsSignals2,HiggsTools}.\\
\addlinespace[2pt]
Flavor & Rare $B\to X_s\gamma$, $B_{s,d}\to\mu^+\mu^-$ and $B\to\tau\nu$ decays, and charged-lepton flavor violation including MEG~II~\cite{MEGII2025}.\\
\addlinespace[2pt]
Collider searches for SUSY & ATLAS and CMS searches for jets, missing transverse momentum and $\tau$ final states. No exclusion in the searches tested~\cite{ATLASMonojet2021,CMS19006,ATLASTau2024,CMSSoftTau2020}.\\\bottomrule
\end{tabular}
\end{table}

\prlsection{Experimental constraints}
Table~\ref{tab:constraints} summarizes the experimental constraints. The benchmarks reproduce the selected neutrino-oscillation intervals and satisfy the relic-density and Higgs requirements. They are compatible with the solar comparisons for the reference response and with the Fermi-LAT dwarf-galaxy limits, for which the model-to-limit ratios are $0.607$, $0.713$ and $0.736$~\cite{FermiDwarfs14}. The collider searches considered do not exclude B1--B3 (\smref{collider}{Sec.~S7}). B1 and B3 yield prompt $\tau$ final states, whereas B2 produces soft chargino decay products. These signatures motivate $\tau$ plus missing transverse momentum and initial-state-radiation searches at the high-luminosity LHC (HL-LHC)~\cite{HLLHCBSM2019}. Their sensitivity to the benchmarks requires a dedicated study.

\prlsection{Conclusions}
We studied an inelastic sneutrino origin of the LZ high-energy event in the inverse-seesaw NMSSM where approximate lepton-number conservation protects the small mass splitting while singlet interactions separate freeze-out from solar-neutrino production.
 Our benchmarks retain the solar capture rate implied by LZ while suppressing neutrino yields through singlet final states. Dalitz conversions strengthen solar constraints, and accompanying gamma-ray boxes probe the same annihilation process through endpoints fixed by the sneutrino and mediator masses. Two benchmarks lie within the projected CTAO sensitivity under the adopted halo and detector assumptions. A compatible gamma-ray signal would independently support such a sneutrino origin of the LZ candidate.


\begin{acknowledgments}
\textit{Acknowledgments}---We thank Yang Zhang and Pengxuan Zhu for helpful discussions. This work is supported by the National Natural Science Foundation of China under Grant No. 12335005, the Natural Science Foundation of Henan Province under Grant No. 252300421771, the Henan Postdoctoral Research Project under Grant No. HN2026054, and the PI Research Fund from Henan Normal University under Grant No. 5101029470335.
\end{acknowledgments}
\prlsection{Data availability} The benchmark inputs, spectra and numerical results are provided with the manuscript. The public experimental inputs are identified in the cited releases.
%

\clearpage
\onecolumngrid
\makeatletter
\renewcommand*{\HyperDestNameFilter}[1]{supp-#1}
\input{size10.clo}
\newgeometry{margin=0.9in}
\global\@colht\textheight
\global\@colroom\textheight
\global\vsize\textheight
\global\hsize\textwidth
\global\linewidth\textwidth
\global\columnwidth\textwidth
\setcounter{page}{1}
\setcounter{section}{0}
\setcounter{secnumdepth}{3}
\setcounter{subsection}{0}
\setcounter{equation}{0}
\setcounter{figure}{0}
\setcounter{table}{0}
\setcounter{footnote}{0}
\renewcommand{\thesection}{S\arabic{section}}
\renewcommand{\theequation}{S\arabic{equation}}
\renewcommand{\thetable}{S\arabic{table}}
\renewcommand{\thefigure}{S\arabic{figure}}
\renewcommand{\theHsection}{S\arabic{section}}
\renewcommand{\theHequation}{S\arabic{equation}}
\renewcommand{\theHtable}{S\arabic{table}}
\renewcommand{\theHfigure}{S\arabic{figure}}
\renewcommand{\section}{\@startsection{section}{1}{\z@}%
  {-3.5ex\@plus -1ex\@minus -.2ex}{2.3ex\@plus .2ex}%
  {\normalfont\Large\bfseries}}
\renewcommand{\subsection}{\@startsection{subsection}{2}{\z@}%
  {-3.25ex\@plus -1ex\@minus -.2ex}{1.5ex\@plus .2ex}%
  {\normalfont\large\bfseries}}
\def\@seccntformat#1{\csname the#1\endcsname\quad}
\let\@hangfrom@section\@hang@from
\let\@hangfroms@section\@hang@froms
\def\figurename{Figure}
\def\tablename{Table}
\setlength{\abovecaptionskip}{10pt}
\setlength{\belowcaptionskip}{0pt}
\long\def\@makecaption#1#2{%
  \vskip\abovecaptionskip
  \sbox\@tempboxa{#1: #2}%
  \ifdim\wd\@tempboxa>\hsize #1: #2\par
  \else\global\@minipagefalse\hb@xt@\hsize{\hfil\box\@tempboxa\hfil}\fi
  \vskip\belowcaptionskip}
\let\table@hook\@empty
\let\@xfloat@prep\@empty
\setlength{\arraycolsep}{5pt}
\setlength{\tabcolsep}{6pt}
\setlength{\parindent}{15pt}
\setlength{\parskip}{2pt}
\setlength{\emergencystretch}{1.5em}
\setlength{\textfloatsep}{14pt}
\setlength{\floatsep}{10pt}
\renewcommand{\floatpagefraction}{0.8}
\renewcommand{\topfraction}{0.9}
\renewcommand{\textfraction}{0.1}
\def\ps@plain{\let\@mkboth\@gobbletwo
  \let\@oddhead\@empty\let\@evenhead\@empty
  \def\@oddfoot{\reset@font\hfil\thepage\hfil}\let\@evenfoot\@oddfoot}
\pagestyle{plain}
\raggedbottom
\microtypesetup{activate=true}
\let\thebibliography\NAT@thebibliography
\let\endthebibliography\endNAT@thebibliography
\let\NAT@bibsetnum\NATx@bibsetnum
\let\NAT@bibsetup\NATx@bibsetup
\def\bibnumfmt#1{[#1]}
\def\refname{References}
\def\bibsection{\section*{\refname}}
\def\bibfont{}
\let\@FMN@list\@empty
\normalsize
\makeatother
\null\vskip 2em
\begin{center}
{\LARGE Supplemental Material: Explaining the LZ High-Energy Recoil Event with Inelastic Sneutrino Dark Matter in Supersymmetry\par}
\vskip 1.5em
{\large\lineskip .5em
\begin{tabular}[t]{c}
Jingwei Lian\textsuperscript{1,2} and Jin Min Yang\textsuperscript{2,3}\\[0.5ex]
{\small\textsuperscript{1}\textit{Henan Institute of Science and Technology, Xinxiang 453003, P. R. China}}\\
{\small\textsuperscript{2}\textit{Centre for Theoretical Physics, Henan Normal University, Xinxiang 453007, P. R. China}}\\
{\small\textsuperscript{3}\textit{Institute of Theoretical Physics, Chinese Academy of Sciences, Beijing 100190, P. R. China}}
\end{tabular}\par}
\vskip 1em
\end{center}
\par
\thispagestyle{empty}

This supplement gives the model inputs and analysis methods. Each benchmark uses a common spectrum fitted to neutrino data.

\hypertarget{sm-model}{}\section{Model and benchmarks}

The ISS-NMSSM adds three pairs of gauge-singlet superfields $\hat N_i,\hat X_i$ to the NMSSM, with lepton numbers $L(\hat N)=-1$ and $L(\hat X)=+1$~\cite{supp:Cao2017,supp:Cao2020,supp:Cao2021gminus2,supp:Cao2020gminus2,supp:Kang2016}. The superpotential is
\begin{equation}
\begin{split}
 W={}&W_{\rm Yukawa}+\lambda\hat S\hat H_u\!\cdot\!\hat H_d+\frac{\kappa}{3}\hat S^3
 +(Y_\nu)_{ij}\hat L_i\!\cdot\!\hat H_u\hat N_j\\
 &+(\lambda_N)_{ij}\hat N_i\hat S\hat X_j+\frac12(\mu_X)_{ij}\hat X_i\hat X_j ,
\end{split}
\end{equation}
where $W_{\rm Yukawa}$ contains the usual quark and charged-lepton Yukawa interactions, and $i,j=1,2,3$. The neutrino and Higgs soft terms are
\begin{align}
 V_{\rm soft}^{\nu}={}&\widetilde L^\dagger m_L^2\widetilde L+
 \widetilde N^\dagger m_N^2\widetilde N+\widetilde X^\dagger m_X^2\widetilde X\nonumber\\
 &+\left[(T_\nu)_{ij}\widetilde L_i\!\cdot\!H_u\widetilde N_j
 +(T_N)_{ij}\widetilde N_iS\widetilde X_j
 +\frac12(b_X)_{ij}\widetilde X_i\widetilde X_j+\hc\right],\\
 V_{\rm soft}^{H}={}&m_{H_u}^2|H_u|^2+m_{H_d}^2|H_d|^2+m_S^2|S|^2
 +\left[\lambda A_\lambda SH_u\!\cdot\!H_d+\frac{\kappa A_\kappa}{3}S^3+\hc\right].
\end{align}
The symmetric matrices $\mu_X$ and $b_X$ have mass dimensions one and two, respectively, and independently violate lepton number by two units. For diagonal trilinear couplings, $T_\nu=Y_\nu A_\nu$ and $T_N=\lambda_N A_N$. All parameters are real, $R$ parity is conserved and sneutrino vacuum expectation values vanish. We use $\sqrt{u^2+d^2}\simeq174\,\GeV$, with $(u,d,s)=(\langle H_u^0\rangle,\langle H_d^0\rangle,\langle S\rangle)$, $\tan\beta=u/d$, $m_D=Y_\nu u$, $M_R=\lambda_Ns$ and $\mu_{\rm eff}=\lambda s$.

\begin{table}[!htbp]
\centering\small
\caption{Inputs and observables supplementing Table~I. Input parameters are specified at $1\,\mathrm{TeV}$, and the listed masses are physical masses. Branching fractions include Dalitz decays, while thermal fractions are lower bounds. The relic abundance is $\Omega_{\snu}h^2$, where $\Omega_{\snu}$ is the present sneutrino density in units of the critical density and $h=H_0/(100\,\mathrm{km\,s^{-1}\,Mpc^{-1}})$. DR2 values use the reference detector response. The IC79 and DR2 statistical prescriptions are defined in Sec.~S5.}
\label{tab:benchmark-summary}
\setlength{\tabcolsep}{7pt}\renewcommand{\arraystretch}{1.02}
\begin{tabular}{lrrr}\toprule
Quantity & B1 & B2 & B3\\\midrule
\multicolumn{4}{l}{\textit{Input parameters}}\\
$\lambda$ & 0.15165 & 0.19938 & 0.43000\\
$\kappa$ & 0.13470 & 0.42782 & 0.87344\\
$\mu_{\rm eff}$ (GeV) & 470.745 & 589.915 & 1375.000\\
$A_\lambda$ (GeV) & 710.113 & 2425.932 & 5674.278\\
$A_\kappa$ (GeV) & -21.670 & -14.478 & -212.605\\
$(Y_\nu)_{33}$ & -0.14500 & 0.08000 & 0.22252\\
$(\lambda_N)_{33}$ & 0.51545 & 0.51338 & 0.96500\\
$(T_\nu)_{33}$ (GeV) & -341.978 & 109.419 & 725.253\\
$(T_N)_{33}$ (GeV) & 1270.299 & 521.595 & 1524.554\\
$(b_X)_{33}$ (GeV$^2$) & -0.197077 & -0.448947 & -0.825112\\
$m_{L_3}$ (GeV) & 519.358 & 659.930 & 1077.798\\
$m_{N_3}$ (GeV) & 1984.787 & 1138.770 & 2164.400\\
$m_{X_3}$ (GeV) & 976.299 & 1337.127 & 2672.447\\
\midrule\multicolumn{4}{l}{\textit{Mass spectrum (GeV)}}\\
$m_{a_s}$ & 74.974 & 76.449 & 64.000\\
$m_{h_{125}}$ & 124.662 & 126.021 & 126.426\\
$m_{h_s}$ & 857.250 & 2550.309 & 5907.547\\
$m_{\snu_1^R}$ & 405.000030 & 600.943494 & 1099.999807\\
$m_{\snu_1^I}$ & 405.000334 & 600.943823 & 1100.000138\\
$m_{\snu_2^R}$ & 425.892567 & 601.200455 & 1122.775044\\
$m_{\snu_2^I}$ & 425.892580 & 601.200475 & 1122.775061\\
$m_{\widetilde\chi_1^0}$ & 480.143 & 601.267 & 1396.817\\
$m_{\widetilde\chi_2^0}$ & 483.249 & 603.933 & 1400.711\\
$m_{\widetilde\chi_3^0}$ & 842.821 & 2551.462 & 2979.783\\
$m_{\widetilde\chi_1^\pm}$ & 482.026 & 602.709 & 1398.648\\
\midrule\multicolumn{4}{l}{\textit{Branching fractions}}\\
$\operatorname{BR}(\snu\snu\to a_sa_s)$ at $v/c=10^{-5}$ & 0.97292 & 0.99084 & 0.97885\\
$\operatorname{BR}(a_s\to\gamma\gamma)$ & 0.91269 & 0.91347 & 0.90989\\
\midrule\multicolumn{4}{l}{\textit{Dark-matter observables and indirect tests}}\\
$\Omega_{\snu}h^2$ & 0.11470 & 0.11159 & 0.11022\\
$\sigma_p^{\rm SI}$ ($10^{-49}\,\mathrm{cm}^2$) & 0.1894 & 1.1489 & 16.4864\\
$\langle\sigma v\rangle_0$ ($10^{-26}\,\mathrm{cm^3\,s^{-1}}$) & 1.5082 & 3.0264 & 9.3710\\
Thermal $a_sa_s$ fraction ($\geq$) & 0.68403 & 0.02624 & 0.87563\\
$C_\odot$ ($10^{21}\,\mathrm{s}^{-1}$) & 4.1663 & 3.4041 & 0.8905\\
IC79 $\Gamma_{\rm ann}/\Gamma_{79}^{90}$ & 0.137 & 0.085 & 0.063\\
IceTracks-DR2 $p^{\rm up}$ & 0.119 & 0.224 & 0.276\\
Fermi-LAT maximum model/limit & 0.607 & 0.713 & 0.736\\
CTAO $\alpha_{95}$ (1.5 kpc core, 500 h) & 0.849 & 1.880 & 6.201\\
\bottomrule
\end{tabular}
\end{table}

In the basis $z=(\snu_{L\alpha},\widetilde N_i^*,\widetilde X_i)^T=(\Phi_R+i\Phi_I)/\sqrt2$, all fields have the same lepton number and each CP sector contains nine real scalars. The indices on $\snu_j^{R,I}$ order states by mass within each sector. In a one-generation example, the quadratic potential is $z^\dagger\mathcal M_0^2z+\tfrac12(z^T\mathcal Bz+\hc)$, where
\begin{align}
 \mathcal M_0^2&=\begin{pmatrix}
 m_L^2+m_D^2+D_L&a_\nu&m_DM_R\\
 a_\nu&m_N^2+m_D^2+M_R^2&a_N\\
 m_DM_R&a_N&m_X^2+M_R^2+\mu_X^2
 \end{pmatrix},\\
 \mathcal B&=\begin{pmatrix}0&0&0\\0&0&M_R\mu_X\\0&M_R\mu_X&b_X\end{pmatrix},
 \qquad \mathcal M_{R,I}^2=\mathcal M_0^2\pm\mathcal B .
\end{align}
Here $D_L=m_Z^2\cos2\beta/2$, $a_\nu=T_\nu u-\mu_{\rm eff}m_D\cot\beta$ and $a_N=T_Ns+\lambda_N(\kappa s^2-\lambda ud)$. For the normalized light eigenvector $\mathbf v=(V_L,V_N,V_X)^T$ of mass $m_0$ at $\mu_X=b_X=0$, define $\Delta_B=\mathbf v^T\mathcal B\mathbf v=2M_R\mu_XV_NV_X+b_XV_X^2$. Nondegenerate first-order perturbation theory gives $m_{R,I}^2\simeq m_0^2\pm\Delta_B$ and hence $\delta\simeq2|\Delta_B|/(m_I+m_R)\simeq|\Delta_B|/m_0$, with the CP-even state lighter for $\Delta_B<0$. For $\mu_X=b_X=0$, lepton number and CP degeneracy are restored. The same breaking terms induce neutrino Majorana masses through scalar loops~\cite{supp:GrossmanHaber,supp:SneutrinoLNV,supp:Hirsch1997BL}. The orthogonal matrices $V^{R,I}$, defined by $\snu_j^{R,I}=\sum_a V^{R,I}_{ja}\Phi_{R,I,a}$, determine $k_Z$. Their light-state rows coincide with $\mathbf v^T$ up to signs in the one-generation unbroken limit, but generally differ after breaking.

The neutralinos $\widetilde\chi_i^0$ ($i=1,\ldots,5$, ordered by increasing mass) are Majorana mass eigenstates obtained by mixing the bino, neutral wino, two neutral Higgsinos and singlino~\cite{supp:NMSSMReview}. The charged winos and Higgsinos form two chargino mass eigenstates $\widetilde\chi_j^\pm$ ($j=1,2$), also ordered by mass. The Higgs sector consists of two doublets, $H_u,H_d$, and a singlet, $S$. After electroweak symmetry breaking, its physical spectrum contains three CP-even scalars $h_i$ ($i=1,2,3$), two CP-odd scalars $a_i$ ($i=1,2$) and a charged pair $H^\pm$. The neutral states mix the corresponding doublet and singlet components. We denote the predominantly singlet CP-even and CP-odd states by $h_s$ and $a_s$, respectively, and the observed SM-like Higgs boson by $h_{125}$.

Table~\ref{tab:benchmark-summary} supplements Table~I. All benchmarks use $\tan\beta=8$ and $M_1=M_2=M_3=3\,\mathrm{TeV}$. Squark soft masses in B1 and B2 are $4\,\mathrm{TeV}$ for the first two generations and $2\,\mathrm{TeV}$ for the third, as in effective supersymmetry~\cite{supp:EffectiveSUSY}. B3 uses $2\,\mathrm{TeV}$ throughout. The spectrum calculation includes their Higgs radiative corrections~\cite{supp:NMSSMReview,supp:ISSHiggsLoops}.

We calculate spectra with SARAH~4.15.4 and SPheno~4.0.7, and dark-matter observables with micrOMEGAs~7.1.4 and CalcHEP~\cite{supp:SARAH4,supp:SPheno,supp:SPhenoBSM,supp:micrOMEGAs7,supp:CalcHEP}.

\hypertarget{sm-neutrino}{}\section{Neutrino masses and mixing}
The inverse seesaw attributes small light-neutrino masses to small lepton-number violation. In the $(\nu_L,N,X)$ basis,
\begin{equation}
 \mathcal M_\nu=\begin{pmatrix}0&m_D&0\\m_D^T&0&M_R\\0&M_R^T&\mu_X\end{pmatrix},
 \qquad m_\nu^{\rm tree}\simeq\Theta\mu_X\Theta^T,\quad\Theta=m_D(M_R^T)^{-1}.
\end{equation}
At $\mu_X=0$, the heavy singlets form Dirac pairs and the light neutrinos are massless at tree level. Small $\mu_X$ generates their masses without requiring small Yukawa couplings~\cite{supp:InverseSeesaw,supp:DeRomeriHirsch2012}.

The independent lepton-number-violating term $b_X$ contributes through neutralino--sneutrino loops~\cite{supp:GrossmanHaber,supp:Hirsch1997BL}. We therefore fit all six independent entries of the real symmetric $\mu_X$ to loop-corrected masses and mixing, adopting normal ordering, $m_1=0.005\,\eV$, $\delta_{\rm CP}=\pi$ and the NuFIT~6.0 IC19 intervals~\cite{supp:NuFIT6,supp:NuFIT60manual}. Tables~\ref{tab:neutrino-summary} and \ref{tab:sm-mux} give the results.

For the light-neutrino charged-current mixing matrix, the polar decomposition $N=(\mathbf1-\eta)U$, with $\eta=\mathbf1-\sqrt{NN^\dagger}$, defines the unitary matrix from which the mixing angles are extracted. We impose marginal upper bounds $(1.4\times10^{-3},1.4\times10^{-4},8.9\times10^{-4})$ on the diagonal entries of $\eta$, $(1.2\times10^{-5},8.8\times10^{-4},1.8\times10^{-4})$ on $(|\eta_{e\mu}|,|\eta_{e\tau}|,|\eta_{\mu\tau}|)$, and $\operatorname{tr}\eta<2.1\times10^{-3}$~\cite{supp:Nonunitarity,supp:NonunitaritySummary}. This criterion uses upper limits only. It does not impose the nonzero lower boundary of the preferred $\eta_{ee}$ interval, below which all three benchmarks lie.

\begin{table}[htbp]
\centering
\small
\caption{Loop-corrected neutrino observables for normal ordering and $\delta_{\rm CP}=\pi$. Nonunitarity is parameterized by $\eta=\mathbf1-\sqrt{NN^\dagger}$, with the marginal upper limits specified in the text.}
\label{tab:neutrino-summary}
\begin{tabular}{lrrr}
\toprule
Quantity & B1 & B2 & B3 \\
\midrule
$m_1$ (eV) & 0.00500463 & 0.00499240 & 0.00500388\\
$m_2$ (eV) & 0.00999975 & 0.00997915 & 0.00999667\\
$m_3$ (eV) & 0.05058963 & 0.05056940 & 0.05059445\\
$\Delta m^2_{21}$ ($10^{-5}\,{\rm eV}^2$) & 7.494871 & 7.465944 & 7.489449\\
$\Delta m^2_{31}$ ($10^{-3}\,{\rm eV}^2$) & 2.534265 & 2.532340 & 2.534759\\
$\sin^2\theta_{12}$ & 0.3074552 & 0.3102816 & 0.3071238\\
$\sin^2\theta_{23}$ & 0.5611421 & 0.5616193 & 0.5611696\\
$\sin^2\theta_{13}$ & 0.0219542 & 0.0219821 & 0.0219599\\
$\eta_{ee}$ ($10^{-5}$) & 0.706095 & 1.636511 & 1.405876\\
$\eta_{\mu\mu}$ ($10^{-5}$) & 0.706098 & 1.636456 & 1.405832\\
$\eta_{\tau\tau}$ ($10^{-5}$) & 11.730998 & 3.956719 & 7.260003\\
$\operatorname{tr}\eta$ ($10^{-5}$) & 13.143192 & 7.229686 & 10.071711\\
\bottomrule
\end{tabular}
\end{table}

\begin{table}[htbp]
\centering\small
\caption{Independent upper-triangle entries of the real symmetric input $\mu_X$, in eV. The lower triangle follows by symmetry. The entries refer to the flavor basis.}\label{tab:sm-mux}
\begin{tabular}{lrrr}\toprule
Entry & B1 & B2 & B3\\\midrule
$(\mu_X)_{11}$ & 536.120432&230.789967&268.469520\\
$(\mu_X)_{12}$ & -237.378420&-102.312911&-119.091026\\
$(\mu_X)_{13}$ & 105.071222&-118.531490&-93.849771\\
$(\mu_X)_{22}$ & 2278.008143&980.616310&1141.706603\\
$(\mu_X)_{23}$ & -357.923712&403.799008&319.732420\\
$(\mu_X)_{33}$ & 192.663841&400.343474&419.717562\\
\bottomrule
\end{tabular}
\end{table}

\hypertarget{sm-thermal}{}\section{Thermal annihilation}

\begin{figure}[!ht]
\centering\includegraphics[width=0.95\textwidth]{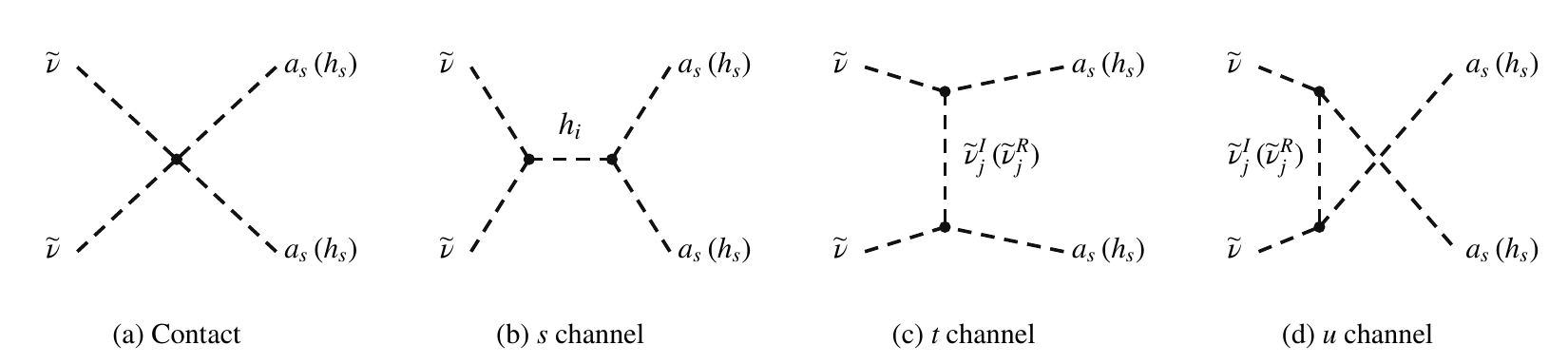}
\caption{Sneutrino annihilation topologies. Labels outside (inside) parentheses denote $a_sa_s$ ($h_sh_s$). The $s$ channel sums over physical CP-even Higgs states. The $t/u$ channels sum over CP-odd sneutrinos for $a_sa_s$ and CP-even sneutrinos for $h_sh_s$. All amplitudes interfere.}\label{fig:sm-topologies}
\end{figure}\FloatBarrier

The thermal average includes all populated odd-particle species,
\begin{equation}
 \langle\sigma_{\rm eff}v\rangle=
 \sum_{ij}\langle\sigma_{ij}v_{ij}\rangle
 \frac{n_i^{\rm eq}n_j^{\rm eq}}{(\sum_k n_k^{\rm eq})^2} .
\end{equation}
The thermal fractions are contributions to the integrated depletion rate, not branching fractions at a single velocity~\cite{supp:GriestSeckel,supp:micrOMEGAs7}. The quoted sums over resolved channels are lower bounds. Fractions classified by initial and final states can overlap and must not be added.

Figure~\ref{fig:sm-topologies} shows the contact and exchange diagrams. All three benchmarks have $m_{h_s}>m_{\snu}$, so annihilation into $h_sh_s$ is kinematically forbidden at low velocities. A lighter CP-even singlet would allow this channel, with doublet mixing relating decays into massive SM particles to Higgs constraints. A singlet-like pseudoscalar instead permits dominant diphoton decays when its doublet component is sufficiently small.

In B1, the thermal distribution can access a resonance above the ground-state pair threshold. Integrating the full interfering amplitudes over $|\sqrt{s}-m_{h_s}|<5\Gamma_{h_s}$ gives at least $63.6\%$ of the depletion rate with a 40-node quadrature. This fraction measures the contribution from the resonance region and differs from the fraction with $a_sa_s$ final states. Kinetic energies of order keV in the cooled solar population are insufficient to reach the resonance.

In B2, electroweakinos dominate thermal depletion, whereas annihilation in the cooled solar population involves the ground-state sneutrinos. The thermal fraction with $a_sa_s$ final states is therefore much smaller than the low-velocity branching fraction. In B3, $\snu\snu$ and $\snu'\snu'$ contribute almost equally at freeze-out, together accounting for at least $87.6\%$ of depletion through $a_sa_s$. Ground-state annihilation into $W^+W^-$ and $ZZ$ remains kinematically allowed even though on-shell electroweak decays of $a_s$ are forbidden. These direct annihilation channels and mediator conversion decays determine the high-energy solar-neutrino spectrum.

\clearpage
\hypertarget{sm-recoil}{}\section{Nuclear recoils}\label{sec:sm-recoil}

\begin{figure}[!ht]
\centering\includegraphics[width=0.70\textwidth]{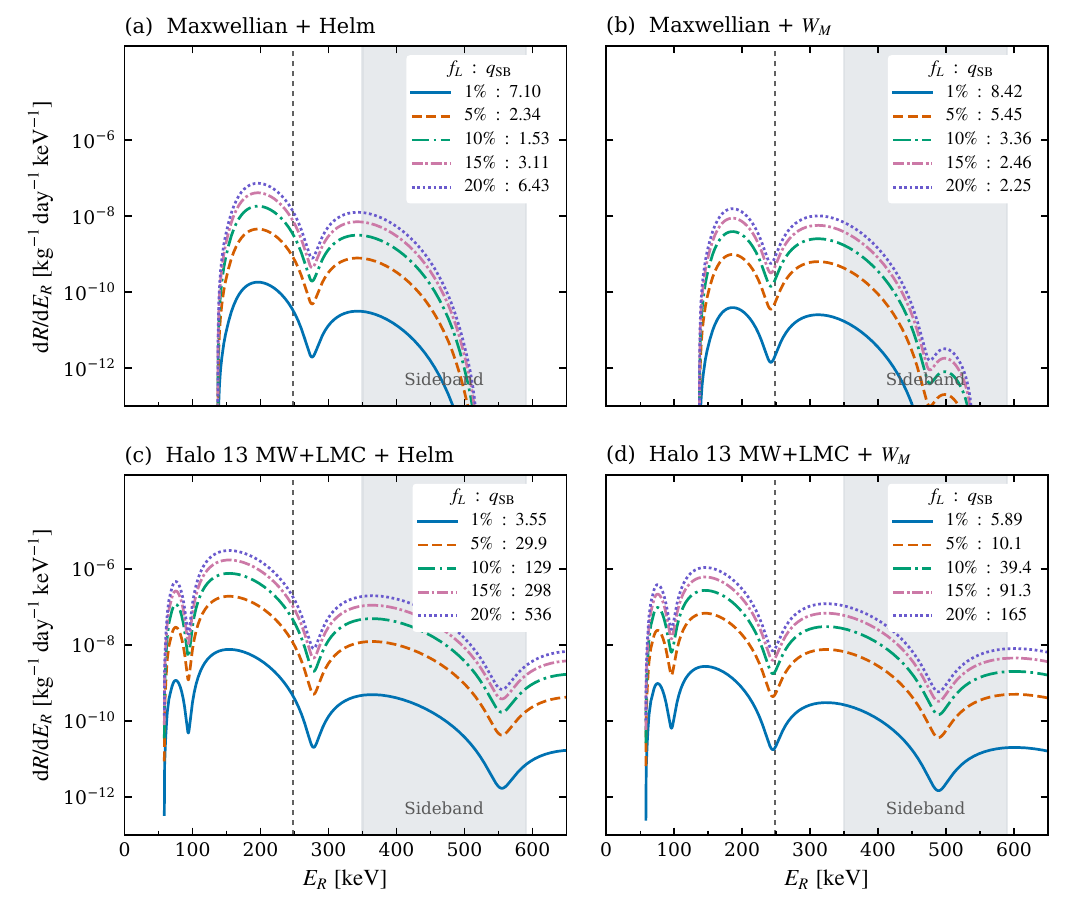}
\caption{True nuclear-recoil spectra in natural xenon for $m_{\snu_1}=600\,\GeV$, $\delta=330\,\keV$ and $k_Z=f_L$, using Helm or $W_M$ responses and the Maxwellian or Halo~13 MW+LMC velocity distribution. Colors and line styles denote $f_L=1\%,5\%,10\%,15\%,20\%$. Legends give $q_{\rm SB}$ after applying the detector response to the signal region. Shading marks the $350$--$590\,\keV$ sideband, and the vertical line marks $248\,\keV$.}\label{fig:sm-genericrecoil}
\end{figure}
\FloatBarrier

For natural xenon, let $\xi_A$ be the mass fraction of isotope $A$ and $Q_{W,A}=(A-Z)-(1-4s_W^2)Z$ its weak charge, with $s_W=\sin\theta_W$. The coherent inelastic recoil rate is
\begin{equation}
 \frac{\dd R_Z}{\dd E_R}=\frac{\rho_{\snu}\sigma_n^{Z,0}}{2m_{\snu}\mu_n^2}
 \sum_A\xi_AQ_{W,A}^2F_A^2\bigl(\sqrt{2m_AE_R}\bigr)
 \eta_v\bigl(v_{\min,A}\bigr),\qquad
 \sigma_n^{Z,0}=\frac{G_F^2\mu_n^2}{2\pi}k_Z^2 .
\end{equation}
Here $\eta_v=\int_{v>v_{\min}}f_{\rm lab}(\bm v)\dd^3v/v$. The superscript 0 denotes the normalization for vanishing mass splitting, while the recoil rate includes the inelastic threshold. Benchmark predictions use Helm form factors and a truncated Maxwellian distribution with $\rho_{\snu}=0.4\,\GeV\,\mathrm{cm}^{-3}$, $v_0=220\,\mathrm{km\,s^{-1}}$, $v_{\rm esc}=544\,\mathrm{km\,s^{-1}}$ and a laboratory-frame boost of $254\,\mathrm{km\,s^{-1}}$.

Elastic Higgs exchange contributes a nuclear amplitude $Zf_p+(A-Z)f_n$, including the relative signs of the nucleon couplings. For $\mathcal L\supset-\tfrac12g_{h_i\snu\snu}h_i\snu^2-g_{h_iNN}h_i\overline NN$,
\begin{equation}
 \sigma_N^{\rm SI}=\frac{\mu_N^2}{4\pi m_{\snu}^2}
 \left|\sum_i\frac{g_{h_i\snu\snu}g_{h_iNN}}{m_{h_i}^2}\right|^2,
 \qquad N=p,n .
\end{equation}
The singlet admixture and interference among scalar-exchange amplitudes can suppress this cross section independently of $k_Z$. Both elastic and inelastic contributions are folded with the published efficiency and a Gaussian energy resolution of $32.53\,\keV$ over true recoil energies of $0.1$--$300\,\keV$, using an exposure of $2.84$ tonne-years and reconstructed energies of $50$--$300\,\keV$~\cite{supp:LZHE,supp:HiggsinoHalo}. The energy-only likelihood is
\begin{equation*}
 \mathcal L=e^{-(s+b)}[g(E_*)+bf_b]G_b(b),\qquad
 f_b=(250\,\keV)^{-1},\qquad
 G_b(b)=\exp\!\left[-\frac{(b-b_0)^2}{2\sigma_b^2}\right],
\end{equation*}
Here $g(E)$, $s$ and $E_*$ follow the Letter. The constraint $(b_0,\sigma_b)=(0.0106,0.0008)$ uses the local background expectation near the candidate as an effective $50$--$300\,\keV$ normalization~\cite{supp:LZHE,supp:HiggsinoHalo}. Profiling $b\ge0$, we fix $-\ln\mathcal L_{\rm ref}=5.698836$ to the pure-$Z$ signal-region best fit over $(m_{\snu}/\GeV,\delta/\keV,|k_Z|)\in[100,1500]\times[100,450]\times[0,0.499]$, with energies in keV and no sideband penalty. The term $2N_{\rm SB}=-2\ln\mathrm{Pois}(0\mid N_{\rm SB})$ assumes unit acceptance and neglects energy migration in the $350$--$590\,\keV$ true-recoil sideband~\cite{supp:Rodd2026}. Elastic scattering contributes $0.084\%$, $0.663\%$ and $2.849\%$ of accepted events for B1--B3.

\hypertarget{sm-halo}{}
Figure~\ref{fig:sm-genericrecoil} compares Helm and zero-momentum-normalized $W_M$ responses from WIMpy~\cite{supp:WIMpyCode}, retaining Helm for $^{124,126}$Xe. At the same local density, we compare the Maxwellian halo with the present-day Halo~13 MW+LMC mean for the best-fit solar position~\cite{supp:SmithOrlikLMC}. The latter is reconstructed from Fig.~11 of that work by interpolating $\log\eta_v$ and tapering the high-speed end to zero. Variations of the taper change accepted counts by less than $0.5\%$. This tests reconstruction, not the full halo uncertainty. LMC infall enhances the high-speed population~\cite{supp:BeslaLMC2019}, with substantial simulation dependence~\cite{supp:OHareHighVelocity2026}. Both signal and sideband counts can increase, so the two templates do not define an uncertainty band. Joint solar tests require consistent recoil and capture distributions.

\hypertarget{sm-capture}{}\section{Solar neutrinos}

For solar capture, we use the same halo density and velocity dispersion as in the recoil calculation, a solar speed of $232\,\mathrm{km\,s^{-1}}$, and 95 solar isotope components with abundances based on Ref.~\cite{supp:Asplund2021}. For stationary target nuclei in the optically thin, single-scattering approximation~\cite{supp:GouldCapture,supp:SolarCapture},
\begin{equation}
 C_\odot=\frac{\rho_{\snu}}{m_{\snu}}\sum_A\int4\pi r^2\dd r\int\!\dd u\,
 \frac{f_\odot(u)}{u}\,w\,\Omega_A^-(w,r),\qquad
 w^2=u^2+v_{\rm esc,\odot}^2(r).
\end{equation}
Here $\Omega_A^-(w,r)$ is the scattering rate into bound orbits for target species $A$. Both elastic and inelastic capture are included for B1--B3, with the latter dominant. Their solar signals are evaluated for a fully cooled population in capture--annihilation equilibrium.

Propagation with nuSQuIDS includes solar oscillations, charged-current absorption, neutral-current energy loss and tau regeneration, followed by vacuum-phase averaging~\cite{supp:SolarNuPropagation,supp:nuSQuIDS}. Monochromatic components are folded separately, with the $0.04$-dex width used only for display. Muons and long-lived light hadrons stop in the Sun, while energy losses of taus and heavy hadrons are included~\cite{supp:Pythia83,supp:PPPCnu}.

\hypertarget{sm-decays}{}
Solar-neutrino and gamma-ray predictions share branching fractions including chargino loops, photon--$Z$ interference, tau polarization and hadronic conversions~\cite{supp:Hazma2022,supp:PDGHadronic}. Stopping is applied only in the Sun, and secondary electromagnetic and photonuclear showers are neglected. Conversion decays increase the mixed-flavor $k_\nu$ by $18.7\%$, $51.5\%$ and $34.3\%$ for B1--B3 at unchanged capture rates.

\hypertarget{sm-solar}{}
We analyze IC79~\cite{supp:nulike,supp:IC79Data}, the ten-year solar search~\cite{supp:IceCubeTenYear}, and fourteen years of IceTracks-DR2 data~\cite{supp:IceTracksDR2,supp:IceTracksDR2Data} separately because their exposures overlap. The high- and low-energy selections of the ten-year search have exposures of $10.4$ and $9.3$ yr, respectively. We consider three flavor prescriptions: muon flavor in both selections, equal flavor fractions in both, and muon flavor at high energies with equal flavor fractions at low energies. The same effective area is used for neutrinos and antineutrinos. The $W^+W^-$ reference at the same DM mass is normalized to $1.721\times10^{19}\,\mathrm{s^{-1}}$ at $1110\,\GeV$ following Ref.~\cite{supp:HiggsinoHalo}. Table~I uses log-linear interpolation. Figure~1(a) uses monotone cubic interpolation in logarithmic mass and cross section.

For IC79, \texttt{nulike}~1.0.9 combines event-count, energy and angular likelihoods for all three samples~\cite{supp:nulike,supp:IC79Data}. We use the flux at 1~AU before Earth propagation and retain detector-normalization uncertainties. A common signal scale $\alpha$ defines $\Gamma_{79}^{90}=\alpha_{90}\Gamma_{\rm ann}$ by $-2\ln[\mathcal L(\alpha_{90})/\mathcal L(0)]=2.71$. Table~\ref{tab:benchmark-summary} reports $1/\alpha_{90}$. Coverage has not been calibrated for these spectra.

Our DR2 analysis selects periods with nonnegative solar declination and events within $20^\circ$ of the Sun. It includes propagation through Earth and uses the averaged muon-flavor response above $100\,\GeV$. The expected count $\lambda_i$ in bin $i$ contains the signal template $s_i$, four background templates $b_{ki}$, and the solar-atmospheric neutrino template $h_i$. Their nonnegative amplitudes are profiled in the Poisson likelihood:
\begin{equation}
 \lambda_i(\mu,\boldsymbol\beta,a)=\mu s_i+\sum_{k=1}^{4}\beta_kb_{ki}+ah_i,
 \qquad \mathcal L=\prod_i\operatorname{Pois}(n_i\mid\lambda_i).
\end{equation}
The one-sided test of $\mu=1$ uses $10^4$ pseudoexperiments at each solar-atmospheric normalization $a=0,1,3$. We report the largest Clopper--Pearson upper tail-probability bound, allocating the error probability among response families. All benchmarks have $p^{\rm up}>0.1$ for the reference response. The weaker-response values are $0.403$, $0.492$ and $0.540$. These tests remain conditional on the signal and detector response.

\hypertarget{sm-nonequilibrium}{}\subsection*{Beyond cooled equilibrium}

Table~\ref{tab:nonequilibrium} gives exploratory points with low-velocity annihilation dominated by $W^+W^-$ and $ZZ$. Orbital evolution determines $F_\odot$ and the spatial distribution used for $k_\nu$. Both points satisfy the adopted neutrino, recoil, relic-density and flavor criteria. Higgs and collider constraints remain untested.

\begin{table}[!htbp]
\centering\small
\caption{Exploratory benchmarks NE-C and NE-R, distinct from B1--B3. The ten-year solar ratio $R_\odot$ uses the normalization and flavor prescriptions of the main benchmarks. DR2 $p^{\rm up}$ tests the predicted signal strength $\mu=1$, with rejection for $p^{\rm up}<0.1$. The ratio $R_{\rm DR2}=1/\mu_{90}$, derived on a finite nuisance-parameter grid, is compared with unity. DR2 values use the reference response unless indicated otherwise. CTAO uses the same response and $1.5$ kpc core as the Letter.}
\label{tab:nonequilibrium}
\setlength{\tabcolsep}{9pt}\renewcommand{\arraystretch}{1.04}
\begin{tabular}{lrr}\toprule
Quantity & NE-C & NE-R\\\midrule
$m_{\snu}$ (GeV) & 600.00010 & 405.00643\\
$\delta$ (keV) & 312.062 & 303.734\\
$k_Z$ & 0.03391 & 0.04214\\
$\Omega_{\snu}h^2$ & 0.12256 & 0.11898\\
$q_{\rm SB}$ & 2.2022 & 2.1690\\
$(\sigma v)_{RR}$ at $v_{\rm rel}=10^{-4}c$ ($10^{-28}\,\mathrm{cm^3\,s^{-1}}$) & 1.0627 & 4.8230\\
$\sigma_n^{\rm SI}$ ($10^{-53}\,\mathrm{cm^2}$) & 1.2501 & 2.7804\\
$F_\odot$ & $4.4483\times10^{-4}$ & $5.5197\times10^{-3}$\\
$\Gamma_{\rm ann}$ ($\mathrm{s^{-1}}$) & $3.3270\times10^{17}$ & $1.1831\times10^{19}$\\
Ten-year $R_\odot$ & 0.0257--0.0266 & 0.2402--0.2507\\
DR2 $p^{\rm up}$ & 0.4977 & 0.3241\\
DR2 weaker response $p^{\rm up}$ & 0.6262 & 0.5271\\
DR2 $R_{\rm DR2}$ & 0.0171 & 0.2643\\
CTAO $\alpha_{95}$, $500$ h & $3.35\times10^{-5}$ & $1.47\times10^{-3}$\\\bottomrule
\end{tabular}
\end{table}

Coannihilation accounts for at least $99.956\%$ of NE-C depletion. In NE-R, at least $77.407\%$ comes from $|\sqrt{s}-m_{h_s}|<5\Gamma_{h_s}$ with singlet-pair final states. Orbital evolution includes capture, scattering, escape, $Z$-mediated decay and annihilation. The latter uses $v_{\rm rel}=10^{-3}c$ rather than an orbital velocity average. DR2 propagation includes Earth matter effects.

Both points satisfy DR2 tests for the reference and weaker responses. Stronger responses are untested. Coverage is $90.7$--$99.8\%$ on the sampled nuisance-parameter grid, without a guarantee between points. Orbital convergence, electroweak elastic scattering and detector systematics require further study, and residual singlet cascades lack a dedicated Dalitz calculation. Setting $\sigma_n^{\rm SI}=10^{-51}\,\mathrm{cm^2}$ raises $F_\odot$ to $0.042$ and $0.64$, illustrating sensitivity to elastic scattering rather than a calculated correction. Gamma-ray signals remain below the assumed CTAO sensitivity.

\hypertarget{sm-photons}{}\section{Gamma rays}

Prompt gamma-ray spectra use the same low-velocity annihilation fractions and mediator branching fractions as the solar calculation, with no particle stopping. For the diphoton cascade, the box endpoints are~\cite{supp:GammaBoxes2012}
\hypertarget{sm-box}{}
\begin{equation}
 E_\pm=\frac{m_{\snu}}{2}(1\pm\sqrt{1-r^2}),\qquad r=m_{a_s}/m_{\snu} .
\end{equation}
The effective cross section normalizing an inclusive box template with four photons per annihilation is
\begin{equation}
 \langle\sigma v\rangle_{\rm box}=\langle\sigma v\rangle_0
 \operatorname{BR}(\snu\snu\to a_sa_s)\operatorname{BR}(a_s\to\gamma\gamma).
\end{equation}
The inclusive photon yield contains one power of $\operatorname{BR}(a_s\to\gamma\gamma)$, while its square gives the exclusive four-photon probability. Detector spectra include the box and conversion photons.

We use the public bin-by-bin likelihoods from the 14-year Fermi-LAT dwarf-galaxy analysis over $0.5$--$1000\,\GeV$~\cite{supp:FermiDwarfs14,supp:FermiSEDData}. For dwarf $d$, the predicted prompt flux is
\begin{equation*}
 \frac{\dd\Phi_d}{\dd E}=\frac{\langle\sigma v\rangle_0J_d}{8\pi m_{\snu}^2}\frac{\dd N_\gamma}{\dd E},
\end{equation*}
where $J_d$ is the density-squared integral over the line of sight and source aperture, and $\dd N_\gamma/\dd E$ includes the box and continuum contributions per annihilation. We integrate this spectrum over each energy bin and match either its photon flux or its energy flux to the published reference template. Likelihood interpolation is restricted to the released flux grids.

For each reconstruction, we combine energy bins and dwarfs, profiling each $\log_{10}J_d$ independently with a Gaussian constraint using the published central value and uncertainty. The one-sided $95\%$ upper limit on $\langle\sigma v\rangle_0$ is set by $-2\Delta\ln\mathcal L_{\rm prof}=2.71$ relative to the best fit with nonnegative annihilation cross section. We analyze separately the 30-dwarf sample with measured $J$ factors and the 42-dwarf benchmark sample, which also includes estimated $J$ factors, using both ordinary and weighted likelihood grids. Together with the two flux mappings, these give eight reconstructions. Table~\ref{tab:benchmark-summary} quotes their largest model-to-limit ratio.

For CTAO, the predicted signal count in reconstructed-energy bin $j$ is
\begin{equation}
 S_j=\frac{T\langle\sigma v\rangle}{8\pi m_{\snu}^2}
 \int\!\dd E\,\frac{\dd N_\gamma}{\dd E}
 \int\!\dd\Omega\,J(\Omega)A_{\rm eff}(E,\Omega)P_{\rm ROI}(E,\Omega)
 \int_{E_j}^{E_{j+1}}\!\dd E_{\rm rec}\,D(E_{\rm rec}\mid E,\Omega).
\end{equation}
Here $T$ is the exposure, $\dd N_\gamma/\dd E$ the photon yield per annihilation, $J(\Omega)$ the line-of-sight integral of the squared DM density, $A_{\rm eff}$ the effective area, $P_{\rm ROI}$ the probability of reconstruction within the angular region of interest, and $D$ the energy-dispersion density. We use the azimuth-averaged South prod3b-v2 response at a zenith angle of $20^\circ$, with $500$ h of exposure and event selections optimized for $50$ h~\cite{supp:CTAOProd3b}. The Galactic-center pointing has an unmasked reconstructed aperture of radius $2^\circ$, and the true arrival directions extend to $3.5^\circ$. The Einasto profile has local density $\rho_\odot=0.4\,\GeV\,\mathrm{cm}^{-3}$, solar Galactocentric distance $r_\odot=8.5\,\mathrm{kpc}$, scale radius $r_s=20\,\mathrm{kpc}$ and shape parameter $0.17$. We consider no core or a constant-density core of radius $1.5$ kpc.

We use 60 logarithmic reconstructed-energy bins from $30\,\GeV$ to $30\,\mathrm{TeV}$. Backgrounds include residual cosmic rays, Base MAX diffuse emission and three astrophysical source templates with free nonnegative normalizations~\cite{supp:BaseMAX,supp:BaseMAXData,supp:CTAGCSources}. The combined cosmic-ray and diffuse component has Gaussian priors of width $0.1$ on its logarithmic normalization and spectral tilt, together with independent additive shifts of $2.5\%$ in each bin. Background-only Asimov data with profiled nuisance parameters define the expected upper limit at $q_{95}=2.71$. The energy response smears the box edges and redistributes continuum photons.
Without a core, the CTAO ratios rise to $10.8$, $23.8$ and $79.4$. The approximate H.E.S.S. comparison~\cite{supp:HESSIGS2022} uses $546$ h, $15\%$ energy resolution and a reflected-OFF approximation. A $1.5$ kpc core reduces the ON-minus-OFF contrast to $0.00627$ of the uncored value. The resulting ratios are $0.304$, $0.221$ and $0.341$, versus $48.4$, $35.3$ and $54.4$ without a core.

\hypertarget{sm-constraints}{}\section{Low-energy constraints and collider searches}\label{sec:sm-constraints}

HiggsBounds tests additional-Higgs exclusions, and HiggsSignals evaluates the observed Higgs mass and signal strengths through HiggsTools~\cite{supp:HiggsBounds5,supp:HiggsSignals2,supp:HiggsTools}. With a Higgs-mass theory uncertainty of $3\,\GeV$, B1--B3 satisfy the exclusion tests and give $\Delta\chi^2=3.75$, $2.93$ and $2.69$ relative to the SM, below the chosen threshold of $6.18$. The relic-density interval $0.096\leq\Omega h^2\leq0.144$ allows a $20\%$ theory tolerance around the Planck central value~\cite{supp:Planck2018}. Local signal rates use the full assumed DM density. The LZ2024 SI model-to-limit ratios are $0.0015$, $0.0063$ and $0.0499$~\cite{supp:LZ2024}.

FlavorKit~\cite{supp:FlavorKit} gives $\mathcal B(B\to X_s\gamma)=(3.39,3.30,3.24)\times10^{-4}$ and $\mathcal B(B_s\to\mu^+\mu^-)=(3.18,3.18,3.19)\times10^{-9}$, within the chosen intervals $[3.11,3.87]\times10^{-4}$ and $[2.80,3.88]\times10^{-9}$. The $B_s$ comparison uses the branching-ratio convention of the calculation. B1 and B2 also satisfy the applied $B_d\to\mu^+\mu^-$, $B\to\tau\nu$ and charged-lepton-flavor constraints, including the MEG~II limit on $\mu\to e\gamma$~\cite{supp:MEGII2025}.

\hypertarget{sm-collider}{}
Production cross sections at $13\,\mathrm{TeV}$ are calculated with MadGraph5\_aMC@NLO and NNPDF2.3 LO parton distributions~\cite{supp:MadGraph5,supp:NNPDF23QED}. SModelS~3.2.0 compares electroweak and colored-particle production with ATLAS and CMS limits~\cite{supp:SModelS3}. Among the matched searches, the most restrictive are monojet for B1 and B2, and multijets plus missing transverse momentum for B3~\cite{supp:ATLASMonojet2021,supp:CMS19006}. All corresponding rate-to-limit ratios are below $0.08$.

A separate B1 simulation with Pythia~8 and parameterized efficiencies tests the ATLAS $\tau$ and CMS soft-$\tau$ searches~\cite{supp:Pythia83,supp:ATLASTau2024,supp:CMSSoftTau2020}. Matching-scale and $\tau$-polarization variations, with Monte Carlo and production-rate uncertainties, give an upper ATLAS rate-to-limit estimate of $0.36$. A factor-of-two acceptance variation raises it to $0.72$. The predicted signal yield in the CMS high-$m_T$ selection is below two events. No exclusion is found within the tested topology coverage and acceptance uncertainties.

At the HL-LHC ($14\,\mathrm{TeV}$, $3\,\mathrm{ab}^{-1}$)~\cite{supp:HLLHCBSM2019}, B1 and B3 motivate $\tau$ plus missing transverse momentum searches. Their charginos decay to $\tau\snu$ with branching fractions of $99.8\%$ and $99.2\%$, and mass differences of $77.0$ and $298.6\,\GeV$, respectively. B2 instead motivates searches using initial-state radiation to identify its soft chargino cascade. B1 also permits $h_s(857\,\GeV)\to a_sa_s\to4\gamma$ with a branching fraction of approximately $0.55$, relating the light pseudoscalar to the thermal resonance. Production depends on the small doublet admixture of $h_s$. The sensitivity of these searches requires dedicated studies.

\begingroup\fontsize{9}{10}\selectfont\setlength{\bibsep}{0pt}

\endgroup

\end{document}